\documentclass[aps,prl,twocolumn,superscriptaddress,showpacs]{revtex4}
\usepackage{graphicx}

\begin{document}

\title{The dynamical Casimir effect in a superconducting coplanar waveguide}

\author{J.R.~Johansson}
\affiliation{Advanced Science Institute, The Institute of Physical and Chemical
Research (RIKEN), Wako-shi, Saitama 351-0198, Japan}
\affiliation{Microtechnology and Nanoscience, MC2, Chalmers University of
Technology, SE-412 96 G{\"o}teborg, Sweden}
\author{G.~Johansson}
\affiliation{Microtechnology and Nanoscience, MC2, Chalmers University of
Technology, SE-412 96 G{\"o}teborg, Sweden}
\author{C.M.~Wilson}
\affiliation{Microtechnology and Nanoscience, MC2, Chalmers University of
Technology, SE-412 96 G{\"o}teborg, Sweden}
\author{Franco Nori} 
\affiliation{Advanced Science Institute, The Institute of Physical and Chemical
Research (RIKEN), Wako-shi, Saitama 351-0198, Japan}
\affiliation{Physics Department, The University of Michigan, Ann Arbor, Michigan 48109-1040, USA}

\date{\today}

%
\begin{abstract}

We investigate the dynamical Casimir effect in a coplanar waveguide (CPW) terminated by a superconducting quantum interference device (SQUID). Changing the magnetic flux through the SQUID parametrically modulates the boundary condition of the CPW, and thereby, its effective length. Effective boundary velocities comparable to the speed of light in the CPW result in broadband photon generation which is identical to the one calculated in the dynamical Casimir effect for a single oscillating mirror. We estimate the power of the radiation for realistic parameters and show that it is experimentally feasible to directly detect this nonclassical broadband radiation.

\end{abstract}

\pacs{85.25.Cp, 42.50.Lc, 84.40.Az}

\maketitle

%

Two parallel mirrors in empty space are attracted to each other due to the vacuum fluctuations of the electromagnetic field, because of the different mode density inside compared to outside of the mirrors. This striking effect of quantum electrodynamics (QED) was predicted by Casimir in 1948 and since then it has also been verified experimentally (see, e.g., Ref. \cite{Milonni1994}).

If the mirrors move, there is also a mismatch between vacuum modes at different instances of time. It was predicted \cite{Moore1970} that this may result in the creation of real photons out of vacuum fluctuations. This dynamical Casimir effect (DCE) also holds for a single mirror subject to nonuniform acceleration in empty space \cite{Fulling1976}. Although receiving considerable interest \cite{Kardar1999,Dodonov2001} since its theoretical discovery, there is still no experimental verification of the DCE. This is mainly due to the fact that the rate of photon production is nonnegligible only when the mirror velocity approaches the speed of light, ruling out the use of massive mirrors \cite{Braggio2005}. Proposals for the experimental verification of the DCE have suggested rapidly changing the field boundary conditions in other ways, e.g., 
using lasers to modulate the reflectivity of a thin semiconductor film \cite{Crocce2004,Braggio2005} or the resonance frequency of a superconducting stripline resonator \cite{Segev2007}. 

\begin{figure}[ht]
\includegraphics[width=8.6cm]{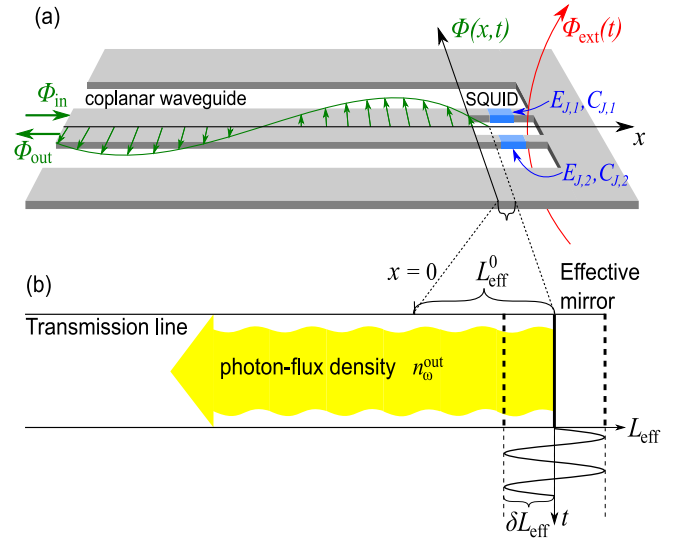}
\caption{(Color online) 
(a) A schematic diagram representing a CPW terminated to ground through a SQUID-loop. The effective inductance of the SQUID can be tuned by the magnetic flux $\Phi_{\rm ext}(t)$, 
providing a tunable boundary condition. (b) The setup in (a) is equivalent to a short-circuited transmission line with a tunable length $L_{\rm eff}$, i.e., with a tunable ``mirror''. We analyze this system using the input/output formalism, which gives the spectrum of the scattered outgoing field, $\Phi_{\rm out}$, as a function of the incoming field, $\Phi_{\rm in}$, in the coplanar waveguide.}
\label{Fig:schematic-device}
\end{figure}

Here we show that a coplanar waveguide (CPW) terminated by a superconducting quantum interference device (SQUID), as shown in Fig.~\ref{Fig:schematic-device}, is a very promising system for experimentally observing the DCE. The inductance of the SQUID can be controlled by the local magnetic flux threading the loop, giving a tunable boundary condition that is equivalent to that of a short-circuited transmission line with a tunable length. Because there is no massive mirror moving, the velocity of the effective boundary can approach the speed of light in the transmission line. The photon production from the vacuum can thus be made experimentally detectable, and its photon spectrum is identical to the one calculated in the DCE for a single oscillating mirror \cite{Lambrecht1996}.

Building on work on superconducting circuits for quantum information \cite{You2005, Wendin2006, Clarke2008}, there has recently been rapid progress in the field of circuit QED, where the interaction between artificial superconducting atoms and the electromagnetic field in microwave cavities is studied. Recent achievements include strong coupling between artificial atoms (qubits) and resonators \cite{CircuitQED}, single-artificial-atom lasing \cite{Astafiev2007}, and Fock-state generation \cite{Hofheinz2008}.  Cavities with tunable frequencies and low dissipation have also been  demonstrated \cite{Palacois2008,Yamamoto2008}, and it has been shown that the resonance frequency can be changed by hundreds of linewidths on a timescale much shorter than the photon lifetime in the cavity \cite{Sandberg2008}. There are also recent theoretical suggestions to observe the DCE in cavity geometries including superconducting qubits \cite{Takashima2008,Dodonov2008}, and also a suggestion to use a CPW where the center conductor is replaced by an array of SQUIDs \cite{Lehnert2008} to simulate the Hawking radiation \cite{Nation2009}.

%
We consider a superconducting CPW with characteristic capacitance $C_0$ and inductance $L_0$ per unit length. The CPW is terminated at $x=0$ through a SQUID-loop threaded by an external flux $\Phi_{\rm ext}(t)$, as shown in Fig.~\ref{Fig:schematic-device}. Since the system contains Josephson junctions, it is convenient to describe the electromagnetic field in the CPW line by its phase field $\Phi(x,t)=\int^t\!dt'\,E(x,t')$, i.e., the time-integral of the electric field $E(x,t)$ \cite{Devoret1995}. The phase field obeys the massless Klein-Gordon equation and, in second quantized form, is ($x<0$):
\begin{eqnarray}
\label{Eq:PhiTQuantizedDecomposed}
\Phi(x,t)
&=&
\sqrt{\frac{\hbar Z_0}{4\pi}}
\int_{0}^{\infty}\!\!\frac{d\omega}{\sqrt{\omega}}
\left(
  a^{\rm in}_\omega e^{-i(-k_\omega x+\omega t)} +
\right.
\nonumber\\
&+&
\left.
  a^{\rm out}_\omega e^{-i(k_\omega x+\omega t)} + \mathrm{H.c}.
\right),
\end{eqnarray}
where the $a^{\rm in\ (out)}_\omega$ operator annihilates a photon with frequency $\omega$ propagating to the right (left) with velocity $v=1/\sqrt{C_0 L_0}$ and wave vector $k_\omega=\omega/v$, and satisfies the commutation relation $[a^{\rm in\ (out)}_\omega, (a^{\rm in\ (out)}_{\omega'})^\dagger] = \delta(\omega-\omega')$. The characteristic impedance of the CPW is $Z_0=\sqrt{L_0/C_0}$.

We first consider a symmetric SQUID, where the two junctions have equal capacitances ($C_{J,1}=C_{J,2}=C/2$) and Josephson energies ($E_{J,1}=E_{J,2}=E_J$), and later comment on the effects of asymmetry. The SQUID effectively behaves as a single junction with a tunable energy
\begin{equation}
E_J(f)=E_J\sqrt{2+2\cos(f)},
\end{equation}
where $f=2\pi\Phi_{\rm ext}(t)/\Phi_0$, and $\Phi_0 = h/2e$. The junction can equivalently be characterized by its tunable (Josephson) inductance $L_J(f)=(\Phi_0/2\pi)^2 / E_J(f)$,
as long as the phase dynamics is slow compared to the plasma frequency $\omega_p(f)=1/\sqrt{CL_J(f)}$, and the SQUID is only weakly excited.

The effective boundary condition for the field imposed by the SQUID can be derived using quantum network theory \cite{Yurke1984}. Starting from the classical Lagrangian for the circuit, the Heisenberg equations of motion are obtained from canonical quantization. For the system under consideration this analysis was performed in Ref.~\cite{WallquistPRB2006}. Here we are now interested in macroscopic SQUID junctions in the phase regime, i.e., when the charging energy is small compared to the Josephson energy, $(2e)^2/2C \ll E_J(f)$, and the quantum fluctuations of the phase across the SQUID are small. In this regime, the boundary condition at $x=0$ becomes
\begin{equation}
\label{Eq:BoundaryCondition}
\frac{(2\pi)^2}{\Phi_0^2}E_J(f)\Phi(0,t)
+
\frac{1}{L_0}
\frac{\partial \Phi(0,t)}{\partial x}
+
C\frac{\partial^2 \Phi(0,t)}{\partial t^2}
= 
0,
\end{equation}
where the last term can be neglected since we are considering dynamics much slower than the plasma frequency of the SQUID. We note that this boundary condition depends parametrically on the tunable effective Josephson energy, $E_J(f)$, providing the tunable boundary condition that will be essential for the remaining discussion.

By inserting the field, Eq.~(\ref{Eq:PhiTQuantizedDecomposed}), in the boundary condition, Eq.~(\ref{Eq:BoundaryCondition}), and performing an integration over time, we find the corresponding boundary condition in frequency space. For a static external flux $f_0$, giving a static Josephson energy $E_J^0 = E_J(f_0)$, we can solve for the output operators $a_\omega^{\rm out}$ in terms of the input operators $a_\omega^{\rm in}$:
\begin{equation}
a_\omega^{\rm out} \ =\ -\;\frac{1+i k_\omega L_{\rm eff}^0}{1- i k_\omega L_{\rm eff}^0}\;a_\omega^{\rm in} \;\equiv\; R(\omega)\;a_\omega^{\rm in},
\end{equation}
where the effective length
\begin{equation}
L_{\rm eff}^0 = L_{\rm eff}(f_0) = \left(\frac{\Phi_0}{2\pi}\right)^2\frac{1}{L_0 E_J^0} = \frac{L_J(f_0)}{L_0} 
\end{equation}
is motivated by comparison to a short-circuited transmission line of length $L$ and its phase factor $-e^{2ikL}$. This effective-length interpretation is valid for $k_\omega L_{\rm eff}^0 \ll 1$, i.e., for frequencies where the SQUID effective length is smaller than the wavelength, or equivalently, $\omega \ll Z_0 C \omega_p^2$, which is satisfied for the parameter regime that we are considering below.

For a time-dependent external flux, resulting in the Josephson energy $E_J[f(t)] = E_J^0 + \delta\!E_J(t)$, we can write the solution in the form,
\begin{eqnarray}
\label{Eq:GeneralAout}
a_\omega^{\rm out} = R(\omega)\,a_\omega^{\rm in}  
- 
\int_{-\infty}^\infty \!\!\!\! d\omega' S(\omega,\omega') \times \nonumber\\
\times \left[\Theta(\omega)(a_{\omega'}^{\rm in}+a_{\omega'}^{\rm out}) + \Theta(-\omega)(a_{\omega'}^{\rm in}+a_{\omega'}^{\rm out})^\dag \right],
\end{eqnarray}
where, 
\begin{eqnarray}
\label{Eq:GeneralS}
S(\omega,\omega') = 
\frac{\sqrt{|\omega/\omega'|}}{2\pi(1-ik_\omega L_{\rm eff}^0)}
\int_{-\infty}^\infty \!\!\!\!dt\;e^{-i(\omega-\omega') t}\;\frac{\delta\!E_J(t)}{E_J^0},
\end{eqnarray}
and where $\Theta$ is the Heaviside step function.

For a small-amplitude harmonic drive $\delta\!E_J(t) = \delta\!E_J \cos(\omega_d t)$, where $\delta\!E_J \ll E_J^0$, the effective length modulation is also harmonic, $L_{\rm eff}(t)=L_{\rm eff}^0+\delta\!L_{\rm eff} \cos(\omega_d t) $, with amplitude $\delta\!L_{\rm eff}=L_{\rm eff}^0 (\delta\!E_J/E_J^0)$. In this case, we can evaluate the integrals in Eqs.~(\ref{Eq:GeneralAout}-\ref{Eq:GeneralS}) and perturbatively solve the resulting equation for the output operators in terms of the input operators,
\begin{eqnarray}
\label{Eq:AoutPerturbative}
a^{\rm out}_\omega
=
R(\omega)\;a^{\rm in}_\omega
+i\frac{\delta\!L_{\rm eff}}{v} \left(\sqrt{\omega(\omega+\omega_d)}
a^{\rm in}_{\omega + \omega_d} + \right. 
\nonumber\\
 \left. \sqrt{|\omega(\omega-\omega_d)|}\left[\Theta(\omega-\omega_d)a_{\omega-\omega_d}^{\rm in} + \Theta(\omega_d-\omega) (a_{\omega_d-\omega}^{\rm in})^\dag \right] \right).
\end{eqnarray}

Considering a large-amplitude harmonic drive, we can expand Eqs.~(\ref{Eq:GeneralAout}-\ref{Eq:GeneralS}) in terms of sideband contributions $\omega_n = \omega+n\omega_d$,
\begin{eqnarray}
\label{Eq:AoutSidebandExpansion}
a^{\rm out}_\omega
=
\sum_{n=-N}^{N} 
c_n\;
\left[
\Theta(\omega_n)
a^{\rm in}_{\omega_n}
+
\;\Theta(-\omega_n)
(a^{\rm in}_{\omega_n})^\dag
\right],
\end{eqnarray}
and numerically solve (for $c_n$) the resulting set of linear equations to arbitrary order $N$, where $N$ is the number of sidebands included in the calculation. Note that we still require both the driving amplitude and the driving frequency to be small compared to the plasma frequency of the SQUID. Using this expansion we have numerically verified that $a^{\rm out}_\omega$ and $(a^{\rm out}_\omega)^\dag$ obtained from Eq.~(\ref{Eq:AoutSidebandExpansion}) satisfies the correct commutation relation $[a^{\rm out}_\omega, (a^{\rm out}_{\omega'})^\dagger] = \delta(\omega-\omega')$, and, similarly, that the perturbative solution in Eq.~(\ref{Eq:AoutPerturbative}) satisfy the same commutation relation to first order in the perturbation parameter $\delta\!E_J$.

Both the perturbative and numerical approaches give, in principle, all properties of the
output field in terms of the input field, and they will be used below to calculate the output
photon-flux density $n^{\rm out}_\omega = \left<(a^{\rm out}_\omega)^\dagger a^{\rm out}_\omega\right>$, as a function of mode-frequency $\omega$, for the thermal input fields  $\bar{n}^{\rm in}_\omega = 1/(\exp(\hbar\omega/k_BT) - 1)$. Using this photon-flux density, the number $N$ of generated photons per second, in a bandwidth $\Delta\omega$, is given by
\begin{equation}
\label{Eq:N}
 N = 
 \frac{1}{2\pi}\int_{\Delta\omega} \!\!d\omega\, n_{\omega}^{\rm out}
 \;\approx\; \frac{\Delta\omega}{2\pi} \;n_{\omega}^{\rm out},
\end{equation}
where the approximation is valid for a small bandwidth $\Delta\omega$, where the relative change in $n_{\omega}^{\rm out}$ is small.

%
For a small-amplitude harmonic drive, we find
\begin{eqnarray}
\label{Eq:nout_w}
n_\omega^{\rm out}
=
\bar{n}^{\rm in}_\omega
&+&
\frac{(\delta\!L_{\rm eff})^2}{v^2}
\;\omega\;|\omega_d-\omega|\;
\bar{n}^{\rm in}_{|\omega_d-\omega|}
\nonumber\\
&+& 
\frac{(\delta\!L_{\rm eff})^2}{v^2}
\;\omega(\omega_d-\omega)\;
\Theta(\omega_d-\omega),
\end{eqnarray}
where we have neglected terms containing the small factor $\bar{n}^{\rm in}_{|\omega_d+\omega|}$, since we are considering $k_BT \ll \omega_d/2\pi$. The output-field photon-flux distribution in Eq.~(\ref{Eq:nout_w}) can be decomposed into three components: The first two are of classical origin; elastically reflected thermal photons (first term) and up-converted thermal photons (second term). The third term is a purely quantum mechanical effect which originates from the vacuum fluctuations. We note that the spectrum of this quantum mechanical radiation is identical to the spectrum of the single-mirror dynamical Casimir effect \cite{Lambrecht1996}.

\begin{figure}[ht]
\includegraphics[width=8.6cm]{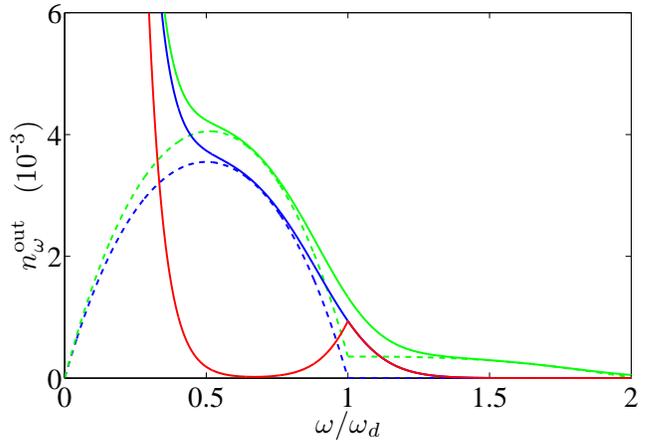}
\caption{(Color online) The photon-flux density as a function of mode-frequency $\omega$. The red curve shows the classical prediction, and the blue (analytical) and green (numerical) curves show the quantum mechanical predictions. The numerical results were calculated using 10 sidebands below and above the center-frequency $\omega$. The discrepancy between the analytical and numerical results is due to the higher-order sideband contributions that is only accounted for in the numerical calculations. Here we have used the parameters $\omega_p/2\pi \approx 36$ GHz, $\omega_d = \omega_p/2$, $\delta\!E_J= E_J^0/4$, $C = 90$ fF, $Z_0\approx55$ $\Omega$, $v \approx 1.2\cdot10^8$ m/s,
and the input-field temperatures 50 mK (solid lines) and zero Kelvin (dashed lines).}
\label{Fig:photon-flux-density}
\end{figure}

The photon-flux-density spectrum of the quantum mechanical radiation has a different frequency dependence compared to that of the reflected thermal photons and the two effects can therefore be clearly distinguished from each other. A signature of the quantum radiation is the parabolic shape in the photon-flux-density spectrum, which has a maximum at $\omega_d/2$, whereas the photon-flux density for the reflected thermal field has maxima at zero frequency and at $\omega_d$, see Fig.~\ref{Fig:photon-flux-density}. Furthermore, in this quantum mechanical radiation process, the photons are created in correlated pairs ($\langle a_{\omega_d/2+\omega}^{\rm out} a_{\omega_d/2-\omega}^{\rm out}\rangle \neq 0$) with frequencies that sum up to the driving frequency, resulting in a squeezing spectrum \cite{Walls1994} with maximum squeezing at $\omega_d/2$.

Using the higher-order expansion in sideband contributions given in Eq.~(\ref{Eq:AoutSidebandExpansion}) we can write the photon-flux density as
\begin{equation}
n^{\rm out}_\omega = \sum_{n=-N}^N |c_n|^2\left[\bar{n}^{\rm in}_{|\omega + n\omega_d|} + \Theta( - \omega - n\omega_d)\right]\;,
\end{equation}
where $c_n$ are numerically-obtained coefficients. Each higher-order sideband gives an additional parabolic contribution, between zero and $n\omega_d$, to the photon-flux-density spectrum. This explains the small discrepancy between the analytical and numerical results in Fig.~\ref{Fig:photon-flux-density}.

\begin{figure}[ht]
\includegraphics[width=8.6cm]{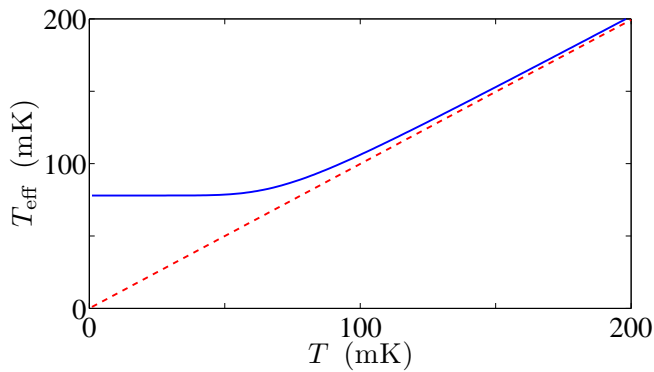}
\caption{(Color online) The effective noise temperature, $T_{\textrm{eff}}$, in the output field of the CPW, versus the input-field temperature, $T$, for purely thermal radiation (red dashed) and for both thermal and quantum radiation (blue solid), at $\omega_d/2$, half the driving frequency. For low enough temperatures, the quantum radiation is significantly larger than the thermal noise level. Here we used the same parameters as in Fig.~\ref{Fig:photon-flux-density}.}
\label{Fig:noise-temperature}
\end{figure}

The experimental verification of the DCE in this system is possible, if for realistic circuit parameters the quantum contribution to the photon flux is distinguishable from the classical thermal contribution. Figure \ref{Fig:photon-flux-density} shows the photon-flux spectral density with the thermal and quantum contributions, for a moderate input mode temperature (50 mK) and typical parameters for superconducting electrical circuits \cite{Sandberg2008}. In this case, the DCE contribution to the photon flux is considerably larger than the classical contribution, in a wide frequency range between $\sim \omega_d/2$ and $\omega_d$. The power per unit bandwidth in this frequency range gives an energy comparable to a few mK, which should be compared with the noise temperature of the typical amplifiers of a few K. Using lock-in techniques and long detection times, these power levels are clearly detectable. Plotting the radiation power per unit bandwidth (in temperature units) at half the drive frequency as a function of the input-field temperature (see Fig.~\ref{Fig:noise-temperature}) further illustrates how the quantum radiation dominates over the thermal radiation for sufficiently low temperatures. With these parameters (see the caption of Fig.~\ref{Fig:photon-flux-density}) the crossover takes place around 70 mK, and the photon production rate $N$ in a $\Delta\omega/2\pi=100$ MHz bandwidth around $\omega_d/2$ is $N \sim10^5$ photons per second, see Eq.~(\ref{Eq:N}).

We have so far only considered symmetric SQUID devices. Here we analyze the case where the SQUID junction capacitances ($C_{J,1},C_{J,2}$) and Josephson energies ($E_{J,1},E_{J,2}$) are slightly asymmetric, i.e., 
$C_J^\Sigma \gg \Delta C_J$ and 
$E_J^\Sigma \gg \Delta E_J$,
where
$C_J^\Sigma = C_{J,1}+C_{J,2}$,
$\Delta C_J = C_{J,2}-C_{J,1}$,
$E_J^\Sigma = E_{J,1}+E_{J,2}$, and
$\Delta E_J = E_{J,2}-E_{J,1}$.
%
Asymmetric capacitances give rise to a source term of the form
$\frac{1}{2}\Delta C_J \frac{\Phi_0}{2\pi} \ddot f(t)$ in the boundary condition Eq.~(\ref{Eq:BoundaryCondition}). This source term produces a coherent oscillating response only at the driving frequency (neglecting the sidebands at zero and $2\omega_d$), which corresponds to a sharp additional peak around $\omega_d$ in the photon-flux-density spectrum. The broadband feature below $\omega_d$ is therefore unaffected by small asymmetries in the junction capacitances.
%
An asymmetry in the Josephson energies gives rise to a correction to the effective SQUID Josephson energy, $E_J(f)$, by the factor $(1-2\Delta E_J/E_J^\Sigma)$. For small asymmetries $\Delta E_J \ll E_J^\Sigma$, there will be a negligible reduction of the photon-flux density.

Finally, we note that it would also be possible to study the DCE in a tunable cavity
geometry \cite{Sandberg2008}. However, here the task of clearly separating the DCE from
the classical effect of parametric amplification is much more demanding since the stable states of the system are essentially identical in the classical and quantum cases.

%
In conclusion, we have studied superconducting coplanar waveguides with parametrically modulated boundary conditions and characterized the spectrum of the photons that are generated in this process. The system can be considered as a solid state analogue to quantum optical setups with moving mirrors, known to generate photons from vacuum fluctuations \cite{Moore1970}. In the present setup, we show that a weak harmonic modulation of the boundary condition can result in broadband photon generation, i.e., the dynamical Casimir effect, and we estimate that it is feasible to detect this radiation in realistic experimental circuits.

%
\begin{acknowledgments}

We would like to thank L. Tornberg, V. Shumeiko, S. Ashhab, G.J. Milburn and P. Bertet for helpful discussions. This work was in part supported by the NSA, LPS, ARO, and NSF Grant No. EIA-0130383.

\end{acknowledgments}

%

\end{document}